\documentclass[12pt]{article}
\usepackage{cite}
\newcommand{\ba}{\begin{array}}
\newcommand{\ea}{\end{array}}
\newcommand{\be}{\begin{equation}}
\newcommand{\ee}{\end{equation}}
\newcommand{\bea}{\begin{eqnarray}}
\newcommand{\eea}{\end{eqnarray}}

\newcommand{\la}{\langle}
\newcommand{\ra}{\rangle}

\newcommand{\p}{\partial}

\def\CF{{\mathcal{F}}}
\def\CP{{\mathcal{P}}}
\def\CQ{{\mathcal{Q}}}
\def\IB{\relax\hbox{$\inbar\kern-.3em{\rm B}$}}
\def\IC{\relax\hbox{$\inbar\kern-.3em{\rm C}$}}
\def\ID{\relax\hbox{$\inbar\kern-.3em{\rm D}$}}
\def\IE{\relax\hbox{$\inbar\kern-.3em{\rm E}$}}
\def\IF{\relax\hbox{$\inbar\kern-.3em{\rm F}$}}
\def\IG{\relax\hbox{$\inbar\kern-.3em{\rm G}$}}
\def\IGa{\relax\hbox{${\rm I}\kern-.18em\Gamma$}}
\def\IH{\relax{\rm I\kern-.18em H}}
\def\IK{\relax{\rm I\kern-.18em K}}
\def\IL{\relax{\rm I\kern-.18em L}}
\def\IP{\relax{\rm I\kern-.18em P}}
\def\IR{\relax{\rm I\kern-.18em R}}
\def\IZ{\relax{\rm Z\kern-.5em Z}}
\def\KZ{Knizhnik-Zamolodchikov }

\def\half{\frac{1}{2}}
\def\p{\partial}
\def\f{\frac}

\begin{document}

\begin{titlepage}

\begin{flushright}
hep-th/0205170\\
OUTP-02-24-P \\
May 2002

\end{flushright}

\vskip 2 cm

\begin{center}
{\LARGE Extended multiplet structure in Logarithmic Conformal Field Theories} \vskip
1 cm

{ \large A. Nichols\footnote{a.nichols1@physics.ox.ac.uk} }

\begin{center}
{\em  Theoretical Physics, Department of Physics, Oxford University\\
1 Keble Road, Oxford, OX1 3NP, UK }
\end{center}

\vskip 1 cm

\vskip .5 cm

\begin{abstract}

We use the process of quantum hamiltonian reduction of $SU(2)_k$,
at rational level $k$, to study explicitly the correlators of the
$h_{1,s}$ fields in the $c_{p,q}$ models. We find from direct
calculation of the correlators that we have the possibility of
extra, chiral and non-chiral, multiplet structure in the $h_{1,s}$
operators beyond the `minimal' sector. At the level of the vacuum
null vector $h_{1,2p-1}=(p-1)(q-1)$ we find that there can be two
extra non-chiral fermionic fields. The extra indicial structure
present here permeates throughout the entire theory. In particular
we find we have a chiral triplet of fields at
$h_{1,4p-1}=(2p-1)(2q-1)$. We conjecture that this triplet algebra
may produce a rational extended $c_{p,q}$ model. We also find a
doublet of fields at $h_{1,3p-1}=(\f{3p}{2}-1)(\f{3q}{2}-1)$.
These are chiral fermionic operators if $p$ and $q$ are not both
odd and otherwise parafermionic.

\end{abstract}

\end{center}

\end{titlepage}

\section{Introduction}

The study of conformal invariance in two dimensions has been a fascinating and
productive area of research for the last twenty years \cite{Belavin:1984vu}. There
is an interesting class of conformal field theories (CFTs) called logarithmic
conformal field theories (LCFTs). In these theories the irreducible primary
operators do not close under fusion and indecomposable representations are
inevitably generated \cite{Gurarie:1993xq}. The operators in the theory have scaling
dimensions that are either degenerate or differ by integers. In these cases it is
possible to have a non-trivial Jordan block structure.

LCFTs have emerged in many different areas for example: WZNW models and
gravitational dressing
\cite{Bilal:1994nx,Caux:1997kq,Kogan:1997nd,Kogan:1997cm,Giribet:2001qq,Gaberdiel:2001ny,Nichols:2001du,Kogan:2001nj,Nichols:2001cv},
polymers \cite{Saleur:1992hk,Cardy,Gurarie:1999yx}, disordered systems and the
Quantum Hall effect
\cite{Caux:1996nm,Kogan:1996wk,Maassarani:1996jn,Gurarie:1997dw,CTT,Caux:1998eu,Bhaseen:1999nm,Kogan:1999hz,Gurarie:1999bp,RezaRahimiTabar:2000qr,Bernard:2000vc,Bhaseen:2000bm,Bhaseen:2000mi,Ludwig:2000em,CardyTalk,Kogan:2001ku},
string theory
\cite{Kogan:1996df,Kogan:1996zv,Ellis:1998bv,Ghezelbash:1998rj,Kogan:1999bn,Myung:1999nd,Lewis:1999qv,Nichols:2000mk,Kogan:2000nw,Moghimi-Araghi:2001fg,Bakas:2002qh,Jabbari-Faruji:2002xz},
2d turbulence
\cite{RahimiTabar:1996dh,RahimiTabar:1997nc,Flohr:1996ik,RahimiTabar:1997ki,RahimiTabar:1996si},
multi-colour QCD at low-x \cite{Korchemsky:2001nx}, the Abelian sandpile
model\cite{Mahieu:2001iv,Ruelle:2002jy} and the Seiberg-Witten solution of
${\mathcal N}=2$ SUSY Yang-Mills \cite{Cappelli:1997qf,Flohr:1998ew}. Deformed
LCFTs, Renormalisation group flows and the $c$-theorem were discussed in
\cite{Caux:1996nm,Rahimi-Tabar:1998ph,Mavromatos:1998sa}. The holographic relation
between logarithmic operators and vacuum instability was considered in
\cite{Kogan:1998xm,Lewis:1998fg}. There has also been much interest on LCFTs with a
boundary \cite{Kogan:2000fa,Ishimoto:2001jv,Kawai:2001ur,Bredthauer:2002ct}. For
more about the general structure of LCFT see
\cite{RahimiTabar:1997ub,Rohsiepe:1996qj,Kogan:1997fd,Flohr:2001tj} and references
therein. Introductory lecture notes on LCFT and more references can be found in
\cite{Tabar:2001et,Flohr:2001zs,Gaberdiel:2001tr,Moghimi-Araghi:2002gk}. A general
approach to LCFT via deformations of the operators has been given in
\cite{Fjelstad:2002ei}.

There has also been much work on analysing the general structure and consistency of
such models in particular the $c_{p,1}$ models and the special case of $c_{2,1}=-2$
which is by far the best understood
\cite{Kausch:1995py,Gaberdiel:1998ps,Kausch:2000fu,Flohr:2000mc}. The key to this
understanding is the fact that one may extend to Virasoro algebra by triplets of
chiral $h_{3,1}=2p-1$ fields \cite{Kausch:1991vg}. The resulting algebra is
sufficient is create a rational LCFT, i.e. one having only a finite number of
irreducible and indecomposable representations \cite{Flohr:1995ea,Flohr:1997vc}. The
aim of this paper is to show that this extended algebraic structure generalises in a
simple way to all $c_{p,q}$ models. We shall leave the question of rationality for
future work.

The WZNW model is of great importance in CFT. Correlation functions in such models
obey differential, Knizhnik-Zamolodchikov , equations \cite{Knizhnik:1984nr} coming
from null states in the theory. The solutions to these equations and correlation
functions for the integrable sector of the $SU(2)_k$ model were studied by
\cite{Zamolodchikov:1986bd,Christe:1987cy}. In the case of the integrable
representations these were previously studied in \cite{MST,ST} in which it was found
that the rational solutions to the \KZ equation were in one-one correspondence with
the extensions of the chiral algebra. There is a simple Dotsenko-Fateev integral
representation for solutions but these do not converge in many cases beyond the
integrable representations. In particular in the cases in which logarithms appear we
have to be very careful when analytically continuing the solutions and it is much
easier, and more convincing, to solve the equations directly. We shall make use of
quantum hamiltonian reduction of $SU(2)_k$ WZNW models, at rational level $k$, which
gives a very efficient procedure to directly calculate differential equations for
the $h_{1,s}$ fields in the $c_{p,q}$ models. By examining the correlation functions
in several examples we shall show that there can exist a very simple structure for a
certain subset of the $h_{1,s}$ operators.

We find that there is a single rational solution generated by the
$h_{1,2p-1}=(p-1)(q-1)$ field corresponding to the vacuum null vector of the
irreducible theory. It is well known that decoupling such a null vector gives us a
complete description of the `minimal' $c_{p,q}$ model \cite{Feigin:1992wv}. However
at this conformal weight we find, in addition, two other primary \emph{fermionic}
non-chiral operators. This extra structure permeates the model.

We found that there are triplets of chiral bosonic fields at
$h_{1,4p-1}=(2p-1)(2q-1)$. These are a natural generalisation of
an algebra, generated by the $h_{1,3}=2q-1$ fields, that appears
in the $c_{1,q}$ models. It has been previously conjectured by M.
Flohr \cite{Flohr:1997vc} that these extended $c_{p,q}$ models
should be formally considered as $c_{3p,3q}$ and we conjecture
that the algebra of such $h_{1,4p-1}$ fields may yield rational
extended $c_{p,q}$ models.  We also observed an extra doublet
structure of the $h_{1,3p-1}=(\f{3p}{2}-1)(\f{3q}{2}-1)$ fields.
If $p,q$ are not both odd then these correspond to chiral
fermionic operators otherwise they are parafermionic.

Recently a particular $SU(2)_k$ theory at rational level, namely
$k=-\f{4}{3}$, was studied \cite{Gaberdiel:2001ny} (See also
\cite{Lesage:2002ch} for a study of $SU(2)_{-1/2}$). It was found
that indecomposable representations were created in the fusion of
admissable representations and that the theory was \emph{not}
rational. On hamiltonian reduction the discrete representations of
$SU(2)$ with $2j \in Z^+$, which are different to the admissable
representations, produce $h_{1,s}$ fields in the $c_{2,3}=0$
model. It would be interesting to see if the type of extended
algebras studied in this paper could be used to construct rational
models of $\widehat{SU(2)}$ at fractional level.
\section{Knizhnik-Zamolodchikov equation}
We consider the $\widehat{SU(2)}$ theory at rational level $k$. The OPE of the
affine Kac-Moody currents is given by:
\bea \label{eqn:SU2KM}
J^3(z) J^{\pm}(w) &\sim& \pm \f{J^{\pm}(w)}{z-w} \nonumber \\
J^+(z)J^-(w) &\sim& \f{k}{(z-w)^2}+\f{2 J^3(w)}{z-w} \\
J^3(z)J^3(w) &\sim& \f{k}{2(z-w)^2}\nonumber \eea
We use the standard Sugawara construction for the stress tensor:
\bea \label{eqn:sugawara} T&=&\f{1}{k+2}\left( \half J^+J^- +\half J^-J^+ + J^3J^3
\right) \eea
which yields a theory with central charge:
\bea c=\f{3k}{k+2} \eea
We consider affine Kac-Moody primary operators having the simple behaviour:
\bea J^a(z) \phi_j(w) \sim \f{t^a \phi_j(w)}{z-w} \eea
where $t^a$ is a spin $j$ matrix representation of $SU(2)$.
We also have affine Virasoro null vectors following from (\ref{eqn:sugawara}):
\be |\chi \ra = ( L_{-1} - \frac{1}{k+2} \eta_{ab}J^a_{-1}J^b_0 ) |\phi \ra \ee
Inserting these null vectors into correlation functions of affine Kac-Moody
primaries one can show that they must satisfy a set of partial differential
equations known as Knizhnik-Zamolodchikov equations \cite{Knizhnik:1984nr}:
\be \left[(k+2) \frac{\p}{\p z_i}+\sum_{j\neq i}\frac{\eta_{ab} t^a_i \otimes
t^b_j}{z_i-z_j} \right] \left<\phi_{j_1}(z_1) \cdots \phi_{j_n}(z_n) \right> =0 \ee
\subsection{Auxiliary variables}
It will be convenient to introduce the following representation for the  $SU(2)$
generators \cite{Zamolodchikov:1986bd}:
\bea \label{eqn:repn} J^+=x^2\frac{\p}{\p x}-2jx, ~~~ J^-=-\frac{\p}{\p x}, ~~~
J^3=x\frac{\p}{\p x}-j \eea
There is also a similar algebra in terms of $\bar{x}$ for the antiholomorphic part.
It is easily verified that these obey the global $SU(2)$ algebra.

We introduce primary fields, $\phi_j(x,z)$  of the affine Lie algebra. Then:
\bea
J^+(z) \phi_j(x,w) &\sim&  \f{\left(x^2\frac{\p}{\p x}-2jx \right)\phi_j(x,w) }{z-w}  \nonumber \\
J^-(z) \phi_j(x,w) &\sim&  \f{ -\frac{\p}{\p x} \phi_j(x,w) }{z-w} \\
J^3(z) \phi_j(x,w) &\sim&  \f{\left( x\frac{\p}{\p x}-j \right)\phi_j(x,w) }{z-w}
\nonumber \eea
The fields $\phi_j(x,z)$ are also primary with respect to the Virasoro algebra with
$L_0$ eigenvalue:
\be h=\frac{j(j+1)}{k+2} \ee
The two point functions and three point functions are fully determined using global
$SU(2)$ and conformal transformations:
\be \label{eqn:2pt} \la \phi_{j_1}(x_1,z_1) \phi_{j_2}(x_2,z_2) \ra = A(j_1)
\delta_{j_1 j_2} x_{12}^{2j_1}z_{12}^{-2h} \ee
\bea \label{eqn:3pt}
\la \phi_{j_1}(x_1,z_1) \phi_{j_2}(x_2,z_2) \phi_{j_3}(x_3,z_3) \ra = C(j_1,j_2,j_3)~~ x_{12}^{j_1+j_2-j_3} x_{13}^{j_1+j_3-j_2} x_{23}^{j_2+j_3-j_1} \\
z_{12}^{-h_1-h_2+h_3} z_{13}^{-h_1-h_3+h_2} z_{23}^{-h_2-h_3+h_1} \nonumber \eea
The $C(j_1,j_2,j_3)$ are the structure constants which in principle completely
determine the entire theory.

For the case of the four point correlation functions of $SU(2)$ primaries the form
is determined by global conformal and $SU(2)$ transformations up to a function of
the cross ratios. Our convention is:
\bea \label{eqn:correl} \langle \phi_{j_1}(x_1,z_1) \phi_{j_2}(x_2,z_2)
\phi_{j_3}(x_3,z_3) \phi_{j_4}(x_4,z_4) \rangle
&=&z_{43}^{h_2+h_1-h_4-h_3}z_{42}^{-2h_2}z_{41}^{h_3+h_2-h_4-h_1} \nonumber \\
& & z_{31}^{h_4-h_1-h_2-h_3}x_{43}^{-j_2-j_1+j_4+j_3}x_{42}^{2j_2}  \\
& & x_{41}^{-j_3-j_2+j_4+j_1}x_{31}^{-j_4+j_1+j_2+j_3}~F(x,z) \nonumber \eea
Here the invariant cross ratios are:
\be x=\frac{x_{21}x_{43}}{x_{31}x_{42}} ~~~ z=\frac{z_{21}z_{43}}{z_{31}z_{42}} \ee
For two and three point functions the \KZ equation gives us no new information.
However for the four point function (\ref{eqn:correl}) using the representation
(\ref{eqn:repn}) we find it becomes a partial differential equation for $F(x,z)$:
\be \label{eqn:KZ} (k+2) \frac{\p}{\p z} F(x,z)=\left[ \frac{\CP}{z}+\frac{\CQ}{z-1}
\right] F(x,z) \ee
Explicitly these are:
\bea
\CP \!\!\!\!&=&\!\!-x^2(1-x)\frac{\p^2}{\p x^2}+((-j_1-j_2-j_3+j_4+1)x^2+2j_1x+2j_2x(1-x))\frac{\p}{\p x} \nonumber \\
& & +2j_2(j_1+j_2+j_3-j_4)x-2j_1j_2 \\
\CQ \!\!\!\!&=&\!\!-(1-x)^2x\frac{\p^2}{\p x^2}-((-j_1-j_2-j_3+j_4+1)(1-x)^2+2j_3(1-x)+2j_2x(1-x))\frac{\p}{\p x} \nonumber \\
& & +2j_2(j_1+j_2+j_3-j_4)(1-x)-2j_2j_3 \eea
For a four point correlator involving a discrete representation of
spin $2j \in Z^+$ we can write the general solution to the \KZ
equation as:
\be \label{eqn:xexp} F(x,z)=\tilde{F}_0(z)+x \tilde{F}_1(z)+x^2\tilde{F}_2(z)+
\cdots +x^{2j}\tilde{F}_{2j}(z) \ee
This allows one to reduce the \KZ equation to a linear \emph{ordinary} differential
equation of order $2j+1$.
\section{Hamiltonian reduction}
When we do a quantum hamiltonian reduction of $SU(2)_k$ theories, by imposing the
constraint $J^+ \sim 1$, it is well known
\cite{Drinfeld:1984qv,Polyakov:1988qz,Alekseev:1989ce,Bershadsky:1989mf,Feigin:1990pn}
that we get the $c_{k+2,1}$ models.

The central charge of the reduced theory is precisely that of the $c_{p,q}$ model
with $k+2=\f{p}{q}$ (we will always take $gcd(p,q)=1$):
\bea \label{eqn:CpqKactable}
c_{p,q}&=&1 - \f{6 (p-q)^2}{pq} \nonumber\\
&=& 13-6 \left( k+2+\f{1}{k+2} \right) \\
h_{r,s}&=&\f{(pr-qs)^2-(p-q)^2}{4pq} \eea
If we perform hamiltonian reduction of the discrete
representations of $SU(2)$ with $2j \in Z^+$ we get the conformal
weights of the $h_{1,2j+1}$ fields in the $c_{p,q}$ model:
\bea \label{eqn:reduced} h_{1,2j+1}&=&\f{j(j+1)}{k+2}-j \eea
Here we follow an elegant realisation of this reduction that
allows us to perform this at the level of the correlation
functions
\cite{Furlan:1991by,Furlan:1993mm,Ganchev:1992af,Ganchev:1993ci}.
Here we shall briefly outline the procedure and do not in any way
attempt to justify its origin. The rather surprising equivalence
suggested is that if one takes the limit $x_i=z_i$ in the
$SU(2)_k$ correlators one obtains those of the reduced $c_{k+2,1}$
model. This is indeed extremely strange as $z$ is a physical
coordinate in the plane and $x$ is an artificial coordinate
introduced to describe the $SU(2)$ structure. Such a procedure of
soldering ordinary and isotopic space was originally suggested by
Polyakov \cite{Polyakov:1989dm}. A quick check of the two and
three point functions of $SU(2)_k$ (\ref{eqn:2pt},\ref{eqn:3pt})
does indeed yield the correct form of these correlators in the
reduced theory with the correct conformal weights
(\ref{eqn:reduced}). However a much more non-trivial statement is
that such a simple procedure also gives the correct four-point
functions. If, rather than expanding $F(x,z)$ as a power series in
$x$ as in (\ref{eqn:xexp}), we expand in the alternative basis:
\be F(x,z)=F_0(z)+(x-z) F_1(z)+(x-z)^2 F_2(z)+ \cdots +(x-z)^{2j} F_{2j}(z) \ee
then it was shown \cite{Furlan:1991by,Furlan:1993mm,Ganchev:1992af,Ganchev:1993ci}
that the lowest component $F_0(z)$, which is the only term surviving in the limit $x
\rightarrow z$, obeys the correct differential equation for the field
$h_{1,2j+1}=\f{j(j+1)}{k+2}-j$ in the $c_{k+2,1}$ model. We have explicitly checked
by hand that this works in a few of the lowest order cases.

As we shall be computing conformal blocks we shall disregard
overall numerical factors which only become important when
considering the consistency of the entire theory. However there
are sometimes subtleties in the reduction process
\cite{Petersen:1996xn} when correlators vanish as $x$ approaches
$z$. This can also be seen as an obstacle in reducing the \KZ
equation to an expression in terms of the lowest component
$F_0(z)$. We did not find any such problems in all the examples
studied in this paper.

This approach gives us a very efficient and practical way to study $h_{1,s}$
correlators up to a very high level. We shall discuss several examples of $h_{1,s}$
correlators the $c_{p,q}$ models in which we have found interesting sets of rational
and logarithmic solutions. Everything that we say could presumably also be
reinterpreted in the $\widehat{SU(2)}$ theory, as solutions for $F_0(z)$ lift up to
solutions for the full $F(x,z)$, but we do not attempt this here.
\section{Vacuum null vector and its fermionic partners}
In this section we shall comment on the vacuum null vector and we find new fermionic
partner fields.
\subsection{Vacuum null vector}
It is known that in the $c_{p,q}$ models by studying the vacuum
null vector we can learn everything about the `minimal' sector of
operators with weights $h_{r,s}$ (\ref{eqn:CpqKactable}) with $1
\le r \le q-1,1 \le s \le p-1$ and identifications
$h_{r,s}=h_{q-r,p-s}$\cite{Feigin:1992wv}.

For example the Ising model at $c_{3,4}=\half$ has a vacuum null vector given by:
\bea \label{eqn:nullIsing} {\mathcal N}= 9 \p^4 T +264 ((\p^2 T) T) -186 (\p T \p
T)) -128 (T(TT)) \eea
One can easily check by using the Virasoro algebra:
\bea T(z) T(w) \sim \f{c}{2(z-w)^4} + \f{2 T(w)}{(z-w)^2}+ \f{\p T(w)}{z-w} + \cdots
\eea
and the normal ordering prescription:
\bea (AB)(w) = \f{1}{2 \pi i} \oint_w \f{dz}{z-w} A(z) B(w) \eea
that this null vector is indeed a primary field of conformal weight $6$:
\bea T(z) {\mathcal N}(w) \sim \f{6 {\mathcal N}(w)}{(z-w)^2}+ \f{\p  {\mathcal
N}(w)}{z-w} + \cdots \eea
In the irreducible theory this null vector should be set to zero in all correlation
functions. In particular the zero mode of this must vanish when applied to Virasoro
primary states $\left. | h \right>$. We know:
\bea
L_n \left. | h \right> &=&0 \quad \quad n \ge 1 \nonumber\\
L_0  \left. | h \right> &=& h \left. | h \right> \eea
and therefore one finds:
\bea {\mathcal N}_0 \left. | h \right> = -4h(2h-1)(16h-1) \left. | h \right> =0 \eea
From this one easily finds the solutions $h=0,\f{1}{2}, \f{1}{16}$
which are well known as the conformal weights of the irreducible
operator content of the Ising model. In general imposing the zero
modes of the $h=(p-1)(q-1)$ vacuum null vector gives us a
polynomial of rank $r=\half (p-1)(q-1)$. Solving this gives us
precisely the $r$ primary operators in the 'minimal' $c_{p,q}$
model \cite{Feigin:1992wv}. Furthermore all fusion rules in this
theory can, in principle, be found from such a null vector. More
details can be found in \cite{Gaberdiel:2001tr} and references
therein.

In particular if we wish to go beyond the minimal $c_{p,q}$ model and consider
fields outside the region with $1 \le r \le q-1,1 \le s \le p-1$ we must not
decouple this vacuum null vector. In order to achieve this we would have to
introduce a logarithmic partner for this field. In the case of the well studied
$c_{p,1}$ models the vacuum null vector is at $h=0$ implying, as is well known, that
all these extended models must have a logarithmic partner for the vacuum itself.

Continuing with the example of the Ising model we can calculate the correlator with
four $h_{1,2p-1}=6$ operators and we find:
\bea \label{eqn:Ising4pt} F(z)=\f{1}{z^6(1-z)^6}
\left(2090z^6-6270z^5+10869z^4-11288z^3+10869z^2-6270z+2090 \right) \eea
This conformal block is easily seen to lead to a well behaved correlator invariant
under all crossing symmetries. By analysing the leading singularity as $z
\rightarrow 0$ we deduce that the two point function of these fields must vanish. To
see that this must be true in general consider the OPE of two vacuum null vector
fields of the irreducible theory having conformal weight $h$. This must have the
form (up to normalisation):
\bea {\mathcal N}(z) {\mathcal N}(w) \sim \f{{\mathcal N}(w)}{(z-w)^h} + \cdots \eea
where $\cdots$ stands for other less singular terms. There cannot be other operators
in the more singular terms as these would also be vacuum null vectors, of lower
conformal weight, contradicting the fact that we are considering the vacuum null
vector of the \emph{irreducible} theory.

This is of course confirmed by explicitly calculating the OPE of
(\ref{eqn:nullIsing}) with itself. However the vanishing of the two point function
of ${\mathcal N}$ immediately implies that the four point function must also vanish.
In order to make the four point function non-zero and realise the conformal block
(\ref{eqn:Ising4pt}) we must have one insertion of the logarithmic partner in order
to make the correlator non-vanishing. This has been discussed in the LCFT literature
many times before (see for example \cite{Flohr:2000mc}). We found in all cases
($c_{p,q}$ with $p,q \le 6$) that there is indeed a single rational solution
generated by the $h_{1,2p-1}=(p-1)(q-1)$ field as we expect. However as we shall see
in the next section there was always two extra \emph{non-chiral} states as well.
\subsection{Non-chiral fermionic partners}
In general we found that the differential equation with four $h_{1,2p-1}=(p-1)(q-1)$
operators always admitted solutions of the form:
\bea \label{eqn:vacnullsolns}
F^{(1)}(z)&=&R_1(z) \nonumber\\
F^{(2)}(z)&=&R_1(z) \ln z + R_2(z) \\
F^{(3)}(z)&=&F^{(2)}(1-z)\nonumber \eea
where $R_1(z)$ and $R_2(z)$ are $\emph{rational}$ functions. The
first solution $F^{(1)}(z)$ is the conformal block of the four
point function of the vacuum null vector, with the subtleties
about insertions of a logarithmic partner, that we have just
discussed. We have already commented that, as this a bosonic
field, we expect it to be invariant under all crossing symmetries:
\bea R_1(z)=R_1(1-z) \quad \quad z^{2h} R_1\left( \f{1}{z} \right) =R_1(z) \eea
The set (\ref{eqn:vacnullsolns}) is clearly closed under monodromy transformations
however in order to be closed under crossing symmetries we must have:
\bea z^{2h} R_2\left( \f{1}{z} \right) =-R_2(z) + \alpha R_1(z) \eea
the constant $\alpha$ is arbitrary but we shall always redefine $F^{(2)}(z)$ by
addition of $F^{(1)}(z)$ to set $\alpha$ to zero.

The other solutions, as we shall presently see, correspond to
extra non-chiral \emph{fermionic} operators. To see this
explicitly it is interesting to consider the example of the
$c_{3,2}=0$ model. This is of great importance in the field of
percolation and polymers
\cite{Saleur:1992hk,Cardy:1992cm,Watts:1996yh}. The vacuum null
vector in this case is the stress tensor $T$ itself and imposing
the vanishing of this in correlators gives us just the `minimal'
topological sector. Considering fields beyond this sector we must
create a logarithmic partner for the stress tensor
\cite{Gurarie:1999yx}.

In this model we found solutions for the $h_{1,3}=2$ conformal blocks:
\bea \label{eqn:solsczero}
F^{(1)}(z)&=& \f{z^2-z+1}{z^2 (z-1)^2} \nonumber\\
F^{(2)}(z)&=& F_1(z) \ln(z) - \f{(5z^5-5z^4+12z^3+12z^2-5z+5)}{24 (z-1) z^4} \\
F^{(3)}(z)&=& F_2(1-z)\nonumber \eea
Before continuing to discuss these solutions we should comment on what occurs if one
instead studies the correlators of the $h_{5,1}=2$ field. Then one finds the
\emph{same} rational block $F^{(1)}(z)$ but a slightly different solution for
$R_2(z)$ in (\ref{eqn:vacnullsolns}) namely:
\bea R_2(z)=\f{\left(5 z^5 -5z^4-16z^3-16z^2-5z+5 \right)}{32 (z-1) z^4} \eea
This seems to be universal and the same rational functions always
appear as a subset of both solutions. In \cite{Gurarie:1999yx} a
possible extension of the conformal algebra at $c=0$ by $h=2$
fields was studied in which there was an extra free parameter $b$.
They found that the operators in the $c_{3,2}=0$ Kac-table
realised only two distinct values of the parameter ($b=\f{5}{6}$,
$-\f{5}{8}$). The appearance of these two different solutions for
$R_2(z)$ seems to be related to these results.

The rational solution $F^{(1)}(z)$ forms a well behaved chiral
correlator on its own and corresponds to the vacuum null vector
$T$. It is easy to see that this is the only primary $(2,0)$
operator in the theory as the other solutions in
(\ref{eqn:solsczero}) on their own do not lead to single-valued
correlators. It is also possible to have local $(2,2)$ operators
in the theory. To see what these are we combine these conformal
blocks with their anti-holomorphic components into the full
correlator:
\be G(z,\bar{z})=\sum_{a,b=1}^{3}{U_{a,b} F^{(a)}(x,z)
\overline{F^{(b)}(x,z)}} \ee
To make this single-valued everywhere we find:
\bea \label{eq:Gfull} G(z,\bar{z})&=&U_{1,1} F^{(1)}(z)
\overline{F^{(1)}(z)}
+ U_{1,2} \Bigl[ F^{(1)}(z) \overline{F^{(2)}(z)}
+ F^{(2)}(z) \overline{F^{(1)}(z)} \Bigr] \nonumber \\
& &+ U_{1,3} \Bigl[ F^{(1)}(z) \overline{F^{(3)}(z)}
 + F^{(3)}(z) \overline{F^{(1)}(z)} \Bigr] \eea
As well as the solution corresponding to the stress tensor $F^{(1)}$ we also have
two other solutions which, as we have logarithms present, do not have a diagonal
form.

Now consider the effect of crossing symmetries on these solutions.
Under $1 \leftrightarrow 3$ we have $z \rightarrow 1-z$ and: \be
F^{(1)} \rightarrow F^{(1)} \quad F^{(2)} \rightarrow F^{(3)}
\quad F^{(3)} \rightarrow F^{(2)} \ee
Under $1 \leftrightarrow 4$ we have $z \rightarrow \f{1}{z}$: \be
F^{(1)} \rightarrow z^4 F^{(1)} \quad F^{(2)} \rightarrow -z^4
F^{(2)} \quad F^{(3)} \rightarrow z^4 \left(-i \pi F^{(1)} -
F^{(2)} + F^{(3)} \right) \ee
We immediately see that the other two solutions are \emph{not} invariant under all
crossing symmetries. To indicate these statistics we add extra labels to these
non-chiral operators. From examining the behaviour under the crossing symmetries we
find that these correspond to \emph{non-chiral fermionic} operators
$\Theta^{\pm}(z,\bar{z})$. To get the correct crossing symmetries we must have:
\bea
\langle \Theta^+(z_1,\bar{z_1})  \Theta^-(z_2,\bar{z_2})
\Theta^-(z_3,\bar{z_3}) \Theta^+(z_4,\bar{z_4})\rangle
&=&|z_{13}|^{-8}|z_{24}|^{-8}\Bigl[ F^{(1)}(z) \overline{F^{(2)}(z)}
+ F^{(2)}(z) \overline{F^{(1)}(z)} \Bigr] \nonumber \\
\langle \Theta^+(z_1,\bar{z_1})  \Theta^+(z_2,\bar{z_2})
\Theta^-(z_3,\bar{z_3})
\Theta^-(z_4,\bar{z_4})\rangle&=&|z_{13}|^{-8}|z_{24}|^{-8}\Bigl[
F^{(1)}(z) \overline{F^{(3)}(z)} + F^{(3)}(z)
\overline{F^{(1)}(z)} \Bigr] \nonumber \eea
By expanding these we see:
\bea \left< \Theta^{\alpha}(z_1,\bar{z_1})  \Theta^{\beta}(z_2,\bar{z_2}) \right> =0
\quad \quad \alpha,\beta=\pm \eea
It appears to be a general fact that all fields beyond the minimal sector of $c=0$
theories have vanishing two point functions. It is the non-vanishing of the four
point functions that gives us a non-trivial theory.

It has been conjectured that fermionic partners to the stress tensor in $c=0$
generate a super-algebra with $U(1|1)$ symmetry \cite{Gurarie:1999yx}. As we have
seen these fields are non-chiral and so certainly cannot be generators of an affine
super-algebra.
\section{Triplet solutions}
In general once one considers the fusion of operators from outside
the minimal region of the $c_{p,q}$ models one generates an
infinite number of Virasoro primary fields. However in the
$c_{p,1}$ models this infinite number of fields can be rearranged
into a finite number with respect to a larger algebra - $W(2,
2p-1,2p-1,2p-1)$. The $h=2$ operator is the stress tensor $T$ and
the other fields $h_{3,1}=2p-1$ are a triplet of fields $W^a$ with
an $SO(3)$ symmetry \cite{Kausch:1991vg}.

Although this algebra was originally found by different methods it
is interesting to see how they arise from the rational solutions
for the conformal blocks. For example in the second member of this
series, the well known $c_{1,2}=-2$ model, we find exactly three
rational four point functions for the $h_{1,3}$ or the $h_{7,1}$
fields both with $h=3$. They are:
\bea
F_{3333}^{(1)}(z) &=& \f{1}{(z-1)^6} z^4 \left(6-6z+z^2 \right)\nonumber\\
F_{3333}^{(2)}(z) &=& \f{1}{z^6(z-1)^6} \left( 2-12z+12z^2+50z^3-225z^4+468z^5-588z^6+468z^7-225z^8 \right.\nonumber\\
&&\left.+50z^9+12z^{10}-12z^{11}+2z^{12} \right)\\
F_{3333}^{(3)}(z) &=& F_{3333}^{(1)}(1-z)= \f{1}{z^6} \left( 1-9z^2+16z^3-9z^4+z^6
\right)\nonumber \eea
Note that $F_{3333}^{(2)}(z)$ is the unique solution satisfying:
\bea
F_{3333}^{(2)}(1-z)=F_{3333}^{(2)}(z) \\
F_{3333}^{(2)}(z)= z^{2h} F_{3333}^{(2)}\left( \f{1}{z} \right)\nonumber \eea It
therefore leads to a correlator that is invariant under all exchanges of operators.
$F^{(1)}(z)$ is the unique solution with no poles as $z \rightarrow 0$ whereas
$F^{(3)}(z)$ is the unique solution with no poles as $z \rightarrow 1$. It is easily
seen that these requirements can be met by assuming that the fields are actually a
bosonic triplet of fields $W^a$ with the following correlators:
\bea
\left< W^+(0) W^+(z) W^-(1) W^-(\infty) \right> &=& F_{3333}^{(1)}(z) \nonumber \\
\left< W^3(0) W^3(z) W^3(1) W^3(\infty) \right> &=& F_{3333}^{(2)}(z) \\
\left< W^+(0) W^-(z) W^-(1) W^+(\infty) \right>&=&F_{3333}^{(3)}(z) \nonumber \eea
These fields are well known in $c=-2$ and are indeed a bosonic
triplet as can be verified from a simple free field construction
\cite{Kausch:1995py}. However we see that the arguments leading us
to this relied only on the existence of the three rational
solutions with the stated pole structure and behaviour under
crossing symmetry. We shall always write our functions $F^{(i)}$
in the same notation as in this example allowing us to immediately
write the correlators and deduce the triplet structure.

One therefore suspects that the same is true in general and that a
triplet of rational solutions, that are closed under crossing
symmetry, will lead to a triplet algebra. We have explicitly
checked for $p \le 9$ that there is indeed a triplet of rational
solutions for the operators of dimension $2p-1$ (the generators of
the triplet symmetry) in the $c_{p,1}$ models.

In generalising this discussion it will be useful to note that in
$c_{p,1}$ we have $h_{3,1}=h_{1,4p-1}$. We will find that it is
the fields $h_{1,4p-1}=(2p-1)(2q-1)$ that become the triplet
fields in the general $c_{p,q}$ models.
\subsection{Correlators in the $c_{1,1}$ model}

The $c_{1,1}=1$ model is a rather peculiar case and so we shall
discuss it separately in this section. The $h_{1,s}$ fields come
from the hamiltonian reduction of the $2j \in Z^+$ operators of an
$SU(2)_{-1}$ theory. The fields have weights:
\bea h_{1,2j+1}=h_{2j+1,1}=j(j+1)-j=j^2 \eea
In this case we find the first few fields have dimensions:
$0,\f{1}{4},1, \cdots$. The $j \in Z^+$ fields have integer
dimensions and all correlators of these fields that we studied
were found to be rational functions.

In particular we found that the $h_{1,3}=1$ fields have three
rational solutions behaving exactly as in the $c=-2$ example and
so we deduce:
\bea \label{eqn:c11triplet}
\left< W^+(0) W^+(z) W^-(1) W^-(\infty) \right>&=& \f{z^2}{(1-z)^2} \nonumber\\
\left< W^3(0) W^3(z) W^3(1) W^3(\infty) \right>&=& \f{(1-z+z^2)^2}{z^2(1-z)^2}  \\
\left< W^+(0) W^-(z) W^-(1) W^+(\infty) \right>&=& \f{(1-z)^2}{z^2} \nonumber \eea
These correlators (\ref{eqn:c11triplet}) are exactly those
corresponding to four point functions of affine currents $J^a$
which generate an $SU(2)_1$ Kac-Moody algebra in the extended
$c_{1,1}=1$ model.

It is interesting to examine this from the point of view of
hamiltonian reduction. We start with the $SU(2)_{-1}$ theory with
three Kac-Moody currents and the triplet of $j=1$ fields. Note in
this case there is potential confusion as the extended fields are
triplets of the $SU(2)_{-1}$ algebra (as they have $j=1$)
\emph{and} also have an extended $SO(3)$ triplet index. After
hamiltonian reduction the $SU(2)_{-1}$ structure is lost but the
extended one remains. What is remarkable, in this example, is that
the extended structure after reduction is in fact \emph{itself} an
$SU(2)$ affine Kac-Moody algebra, this time at level $k=1$. As the
$SU(2)_1$ model is one of the very simplest rational CFTs one may
hope by considering the extended triplet algebra in $SU(2)_{-1}$
that this model should be a relativity simple example of a
rational non-unitary CFT. It is not clear if this theory involves
indecomposable representations or not. The four point correlators
for the irreducible representations were all rational functions
but further fusions may yield other representations.
\subsection{Correlators in the $c_{p,q}$ models}

We found for every $c_{p,q}$ model (we tested $p \le 5, q \le 5$) that there was
always exactly three rational solutions for the $h_{1,4p-1}=(2p-1)(2q-1)$ fields.
Rather more non-trivially if one exchanges $p$ and $q$ the differential equations
are of a different order but the same set of three rational solutions solves both of
them. These triplets appeared to always have a bosonic nature under crossing
symmetry.

As we have discussed the $c_{2,1}=c_{1,2}$ case in the previous
section we shall begin with the first new example: the $c_{2,3}=0$
theory. In the $c_{2,3}=0$ model the solutions are given by:
\bea
F^{(1)}&=& \f{1}{(1-z)^{28}} \left((357106464-2856851712 z+10509841628 z^2-23573986436 z^3 \right. \nonumber \\
&&+36044249670 z^4-39790427248 z^5+32773983814 z^6-20529517008 z^7 \\
&&+9880147186 z^8-3667147120 z^9+1048374600 z^{10}-229634210 z^{11} \nonumber\\
&&\left.+38248769 z^{12}-4810728 z^{13}+452625 z^{14}-30294 z^{15}+1122 z^{16})z^{10} \right)  \nonumber\\
F^{(2)}&=& \f{1}{z^{28} (1-z)^{28} } \left( 2244-60588 z+905250 z^2-9621456 z^3+76497538 z^4 \right. \nonumber\\
&&-459268420 z^5+2096749200 z^6-7334294240 z^7+19760294372 z^8 \nonumber\\
&&-41059034016 z^9 +65547967628 z^{10}-79580854496 z^{11}+72088499340 z^{12} \nonumber\\
&&-36330724836 z^{13}-200733901482 z^{14}+2212292459088 z^{15}-14422439940116 z^{16} \nonumber\\
&&+68562493363130 z^{17}-254028569259777 z^{18}+763908934818536 z^{19}\nonumber\\
&&-1917517271406737 z^{20}+4101816418782654 z^{21}-7599053781520630 z^{22}\nonumber\\
&&+12352604911298080 z^{23}-17809256023135980 z^{24}+22972890487011504 z^{25}\nonumber\\
&&-26689578674273868 z^{26}+28044134317298400 z^{27}-26689578674273868 z^{28}\nonumber\\
&&+22972890487011504 z^{29}-17809256023135980 z^{30}+12352604911298080 z^{31}\\
&&-7599053781520630 z^{32}+4101816418782654 z^{33}-1917517271406737 z^{34}\nonumber\\
&&+763908934818536 z^{35}-254028569259777 z^{36}+68562493363130 z^{37}\nonumber\\
&&-14422439940116 z^{38}+2212292459088 z^{39}-200733901482 z^{40}\nonumber\\
&&-36330724836 z^{41}+72088499340 z^{42}-79580854496 z^{43}+65547967628 z^{44}\nonumber\\
&&-41059034016 z^{45}+19760294372 z^{46}-7334294240 z^{47}+2096749200 z^{48}\nonumber\\
&&\left. -459268420 z^{49}+76497538 z^{50}-9621456 z^{51}+905250 z^{52}-60588 z^{53}+2244 z^{54} \right)\nonumber \\
F^{(3)}&=&  \f{1}{z^{28}} \left( (1122+12342 z+132855 z^2+1026528 z^3+5156450 z^4 \right. \nonumber\\
&&+17580680 z^5+42038555 z^6+70854550 z^7+83500300 z^8+70854550 z^9\\
&&+42038555 z^{10}+17580680 z^{11}+5156450 z^{12}+1026528 z^{13}\nonumber\\
&&\left.+132855 z^{14}+12342 z^{15}+1122 z^{16}) (1-z)^{10}\right) \nonumber \eea
where we have again used the same conventions as before in the
labelling of the $F^{(i)}$. Although the detailed form of the
solutions is extremely complicated the structure is very simple.
In particular their behaviour under crossing symmetry is the same
as in the previous $c=-2$ example. Therefore in an exactly
analogous way to the arguments used there we deduce that the
triplet of dimension $h_{1,7}=15$ chiral fields in $c_{2,3}=0$
behave as:
\bea
\left< W^+(0) W^+(z) W^-(1) W^-(\infty) \right> &=& F^{(1)}(z) \nonumber \\
\left< W^3(0) W^3(z) W^3(1) W^3(\infty) \right> &=& F^{(2)}(z) \\
\left< W^+(0) W^-(z) W^-(1) W^+(\infty) \right>&=&F^{(3)}(z)
\nonumber \eea
It is not yet possible to conclude that all these algebras are closed in the sense
of $W$-algebras as it is extremely difficult to read off the operator content from
the rational correlation functions. We shall see that they do indeed appear closed
as the other $h_{1,s}$ operators that could potentially contribute in the singular
terms of the OPE obey Fermi statistics.
\section{Doublet solutions}
The triplet algebra as we have found, but certainly not proved,
appears in all the $c_{p,q}$ models. However we also found that if
$p$ and $q$ were not both odd then there was also a doublet of
rational solutions with fermionic behaviour generated by the
$h_{1,3p-1}=(\f{3p}{2}-1)(\f{3q}{2}-1)$ fields. If $p$ and $q$ are
both odd then this field does not have $2h \in Z^+$ and so cannot
be a local chiral field and is parafermionic \cite{Fateev:1985mm}.

The most familiar example is the $h_{1,2}$ or $h_{5,1}$ fields at
$h=1$ in the $c_{2,1}=-2$ model where we have the rational
solutions:
\bea
F^1(z)&=&1-\f{1}{z^2} \\
F^2(z)&=&1-\f{1}{(1-z)^2} \eea
$F^{(1)}$ is now the unique solution with no poles as $z
\rightarrow 1$ and we have $F^{(2)}(z)=F^{(1)}(1-z)$. By using
similar crossing symmetry arguments as before we see that there
are two chiral fermionic states:
\bea
\left< \Psi^+(0) \Psi^-(z) \Psi^-(1) \Psi^+(\infty) \right> &=& F^{(1)}(z)\\
\left< \Psi^+(0) \Psi^+(z) \Psi^-(1) \Psi^-(\infty) \right> &=& F^{(2)}(z) \nonumber
\eea
These are precisely the $h=1$ symplectic fermion fields $\Psi^{\pm}(z)$ from the
$c=-2$ theory:
\bea \Psi^+(z) \Psi^-(w) \sim \f{1}{(z-w)^2} \eea
We can also repeat the discussion for the next case in the
$c_{p,1}$ series namely $c_{3,1}=-7$ with $h_{2,1}=\f{7}{4}$. We
then have conformal blocks:
\bea
F^{(1)}(z)&=&\f{1}{\sqrt{z(1-z)}} \f{(z-1)^2}{z^3} \left( 2 z^2+3z+2 \right) \\
F^{(2)}(z)&=& F^{(1)}(1-z) \nonumber \eea
We see that both solutions have branch cuts in the complex plane.
They are therefore not generated by a chiral algebra but instead
by a parafermionic one \cite{Fateev:1985mm}. They still have the
same form as before and so the same arguments can be made to
deduce that this field has in fact a doublet nature:
\bea
\left< \Psi^+(0) \Psi^-(z) \Psi^-(1) \Psi^+(\infty) \right> &=& F^{(1)}(z)\\
\left< \Psi^+(0) \Psi^+(z) \Psi^-(1) \Psi^-(\infty) \right> &=&
F^{(2)}(z) \nonumber \eea
In $c_{2,3}=0$ we also find a similar doublet nature for the
$h_{1,5}=7$ operators:
\bea
F^1(z)&=& \f{1 }{z^{12}}
\left((-22 z^9-44 z^8-323 z^7-859 z^6-1302 z^5-1302 z^4-859 z^3 \right. \\
&& \left.-323 z^2-44 z-22)(1-z)\right) \nonumber\\
F^2(z)&=& \f{1}{(1-z)^{12}}
\left( (22 z^9-242 z^8+1467 z^7-6200 z^6+18475 z^5-37854 z^4 \right.\\
&& \left.+51884 z^3-45424 z^2+22950 z-5100) z \right)\nonumber \eea
We have checked many other $c_{p,q}$ models, with $pq \in 2Z^+$,
and always found $2$ rational solutions with a fermionic symmetry
for the $h_{1,3p-1}=(\f{3p}{2}-1)(\f{3q}{2}-1)$ fields.
\section{General structure}
The structure of rational solutions and Bose/Fermi assignments to
operators is very suggestive. It seems that when the operators had
integer conformal weights an odd number of rational solutions
corresponds to bosonic operators and and an even one to fermionic
ones. The cases we studied all fit into the sequence
$h_{1,2np-1}=(np-1)(nq-1)$ having $2n-1$ rational solutions where
$n=1,2,3,\cdots$ for bosonic fields and
$n=\f{3}{2},\f{5}{2},\cdots$ for fermionic fields.

We checked explicitly in many examples that the $h_{1,2np-1}$
fields are the only chiral $h_{1,s}$ fields by searching for
rational solutions of correlation functions of other operators. If
this is true then we immediately see in the singular terms of the
OPE of two $h_{1,4p-1}$ triplet fields we can only create
$h_{1,2p-1}$ and $h_{1,4p-1}$ fields. The next possible bosonic
field is at $h_{1,6p-1} > 2 h_{1,4p-1}$ and so lies beyond the
singular terms in the chiral OPE. Therefore the triplet algebra
must close as a $W$-algebra with the schematic OPE:
\bea h^a_{1,4p-1} \otimes h^b_{1,4p-1} = \delta^{ab} \left[h_{1,2p-1} \right] +
f^{ab}_c \left[h^c_{1,4p-1} \right] \eea
where $\delta^{ab}$ and $f^{ab}_c$ are the metric and structure constants of $SU(2)$
and $[h]$ denotes an operator and all its descendants. Recall that the $h_{1,2p-1}$
field is the vacuum null vector of the irreducible theory and so $[h_{1,2p-1}]=[1]$.

Throughout this paper we have only considered $c_{p,q}$ with $p,q \in Z^+$. The case
of $c_{p,-q}$ with $c > 25$ can also be obtained from hamiltonian reduction of
$SU(2)_k$ with $k+2 <0$. In this case all discrete representations have negative
dimensions however one also observes rational functions for certain correlators.

It is clear that these structures deserve much closer investigation.
\section{The moduli space of CFTs}
In this section we shall analyse the approach to local logarithmic CFTs in two
particular cases. The first is the well known appearance of an indecomposable
representation and the second is a situation in which operators may have extended
indices, We shall analyse these in the well known $c_{2,1}=-2$ model as the operator
content is particularly well known.

The first correlator that we shall analyse is the original one studied by Gurarie
\cite{Gurarie:1993xq} for the $h_{1,2}=-\f{1}{8}$ operators:
\bea \left< \mu(z_1,\bar{z}_1) \mu(z_2,\bar{z}_2) \mu(z_3,\bar{z}_3)
\mu(z_4,\bar{z}_4) \right> = |z_{13} z_{24} |^{1/2} |z(1-z)|^{1/2} G(z,\bar{z}) \eea
We can easily find the conformal blocks and the unique single-valued combination is:
\bea G(z,\bar{z})= \CF(z) \overline{\CF(1-z)} + \CF(1-z) \overline{\CF(z)} \eea
where $\CF(z)$ is the hypergeometric function: $_2F_1\left( \half,\half;1;z
\right)$. This leads to a correlator invariant under all exchanges of operators. It
is well known that this correlator has logarithmic singularities and it is
interesting to see how these emerge as $c \rightarrow -2$.

To analyse this we examine the $c_{k+2,1}$ model when $k$ is
small. The central charge is:
\bea c=13-6 \left( k+2 + \f{1}{k+2} \right) = -2 -\f{9 k}{2} +O(k^2) \eea
The first few operators in the Kac-table have dimensions:
\bea
h_{1,1}&=&0 \nonumber\\
h_{1,2}&=&\f{3}{4(k+2)}-\half = -\f{1}{8} -\f{3 k}{16} + O(k^2) \\
h_{1,3}&=&\f{2}{k+2}-1= -\f{k}{2} + O(k^2)\nonumber \eea
At the point $k=0$ we have $h_{1,3}=h_{1,1}=0$ but for generic values of $k$ there
is no degeneracy in the levels. We can find the general solutions for the four point
function of $h_{1,2}$ operators. The full correlator is given by:
\bea \left< h_{1,2} (z_1,\bar{z}_1) h_{1,2} (z_2,\bar{z}_2) h_{1,2} (z_3,\bar{z}_3)
h_{1,2} (z_4,\bar{z}_4) \right> = |z_{13} z_{24} |^{-4 h} |z|^{\f{2k+1}{k+2}}
|1-z|^{\f{1}{k+2}} G(z,\bar{z}) \nonumber \eea
where:
\bea G(z,\bar{z})=\sum_{i,j=1}^{2} U_{i,j} F_i(z) \overline{F_j(z)} \eea
and the conformal blocks $F_i(z)$ are found by solving the differential equations or
via the Coulomb gas approach. They are:
\bea
F_1(z)&=&  \f{\Gamma \left(\f{k+1}{k+2}\right) \Gamma \left(\f{k+1}{k+2}\right)}{\Gamma \left(\f{2k+2}{k+2}\right) }   {}_2F_1 \left( \f{1}{k+2}, \f{k+1}{k+2};\f{2k+2}{k+2};z \right) \\
F_2(z)&=& z^{\f{-k}{k+2}} \f{\Gamma \left(\f{1-k}{k+2}\right) \Gamma
\left(\f{k+1}{k+2}\right)}{\Gamma \left(\f{2}{k+2}\right) }   {}_2F_1 \left(
\f{1}{k+2},\f{1-k}{k+2};\f{2}{k+2};z \right) \eea
We have included the normalisations so that we can use standard results. The
solutions $F_1(z)$ and $F_2(z)$ are respectively the conformal blocks for the
contributions from the $h_{1,1}$ and $h_{1,3}$ operators respectively as can be seen
from the leading powers of $z$. However we immediately see that these two solutions
become identical in the limit as $k \rightarrow 0$.

The full correlator must of course be single-valued everywhere. Monodromy around
$z=0$ leads to the requirement that:
\bea \label{eqn:monod}
U_{1,2}e^{2 \pi i \f{k}{k+2}}= U_{1,2} \\
U_{2,1}e^{-2 \pi  \f{k}{k+2}}= U_{2,1}\nonumber \eea
Now for the case of generic values of $k$ we have $\f{k}{k+2} \notin Z$ and we
conclude $U_{1,2}=U_{2,1}=0$ and the correlator must be diagonal:
\bea G(z,\bar{z})=U_{1,1} |F_1(z)|^2 + U_{2,2} |F_2(z)|^2 \eea
Now imposing the monodromy around $z=1$ leads to the condition
\cite{Dotsenko:1984nm}:
\bea \f{U_{1,1}}{U_{2,2}}=\f{\sin \pi(a+b+c) \sin \pi b }{\sin \pi a \sin \pi c}
\eea
where $a=\f{-2k-1}{k+2}~,~~b=c=\f{-1}{k+2}$.
Expanding this in the limit $k \rightarrow 0$ we get:
\bea \f{U_{1,1}}{U_{2,2}}=-1 + O(k^2) \eea
Therefore:
\bea \label{eqn:mumumumu} G(z,\bar{z})=U_{1,1} \left( |F_1(z)|^2 - |F_2(z)|^2
\right) \eea
The minus sign is absolutely crucial. It signifies that we have negative norm
states. Logarithms can occur when these are cancelled to leading order by the
positive norm states. Expanding $F_1$ and $F_2$ gives:
\bea
F_1(z) = \pi \CF(z) + k C(z)  \\
F_2(z) = \pi \CF(z) +k D(z) \eea
where:
\bea C-D=\f{\pi^2}{2} \CF(1-z) \eea
and $\CF(z)$ is again the hypergeometric function: $_2F_1\left(
\half,\half;1;z \right)$. In order to make the full correlator
non-vanishing in the limit $k \rightarrow 0$ we will have to
choose the overall normalisation of the four point function
(\ref{eqn:mumumumu}) to be $U_{1,1} \sim \f{1}{k}$. With this
choice we find that we have a smooth limit as $k \rightarrow 0$.
\bea
G(z,\bar{z})&=&\f{1}{k} \left( (\pi \CF +kC)\overline{(\pi\CF + kC )} - (\pi \CF+kD)\overline{( \pi \CF +k D )} \right) \\
&&\rightarrow \left[ \CF(z) \overline{\CF(1-z)} + \CF(1-z) \overline{\CF(z)} \right]
\nonumber \eea
In this case we were able to get a smooth approach to a logarithmic correlator from
a non-logarithmic one. One might be therefore tempted to think that LCFT is merely
some continuous limit of ordinary CFT. However we shall soon see that this is
\emph{not} always the case.

To illustrate this we shall examine the correlator:
\bea
\left< h_{1,2} (z_1,\bar{z}_1) h_{1,2} (z_2,\bar{z}_2) h_{1,3} (z_3,\bar{z}_3) h_{1,3} (z_4,\bar{z}_4) \right> &=&  |z_{34}|^{4h_{1,2}-4h_{1,3}} |z_{24}|^{-4h_{1,2}}|z_{13}|^{-4h_{1,2}} \nonumber\\
&& ~~~ |z|^{\f{2k+1}{k+2}} |1-z|^{\f{2}{k+2}} G(z,\bar{z}) \eea
Evaluating this correlator for the $c=-2$ theory we get two solutions:
\bea
\CF_1(z)&=& (1-z)^{-1/2} \\
\CF_2(z)&=& (1-z)^{-1/2} \arctan( \sqrt{z-1})\nonumber \eea
The function $\arctan( \sqrt{z-1})$ has the following behaviour near $z=0$
\footnote{This is most easily seen using: \bea
\arctan(\sqrt{z-1}) = \int \f{1}{2z\sqrt{z-1}} ~ dz &=& \f{1}{2i} \int \left[ \f{1}{z} + \f{1}{2} + \f{3}{8}z + \cdots \right] dz \nonumber \\
&=& \f{1}{2i} \ln z + {\rm regular} \nonumber \eea}:
\bea \arctan(\sqrt{z-1}) \sim  - \f{i}{2} \ln z + {\rm regular} \eea
Making $G(z,\bar{z})$ single valued requires no logarithmic branch cuts and
therefore we have \emph{two} possible single-valued correlators:
\bea G(z,\bar{z})=U_{1,1}\CF_1 \overline{\CF_1} + U_{1,2} \left( \CF_1
\overline{\CF_2} + \CF_2 \overline{\CF_1} \right) \eea
The solution with $U_{1,2}=0$ corresponds to the correlator:
\bea \left< \mu (z_1,\bar{z}_1) \mu (z_2,\bar{z}_2) \Omega (z_3,\bar{z}_3) \Omega
(z_4,\bar{z}_4) \right> =  |z_{12}|^{1/2} \nonumber \eea
The other solution with logarithmic terms corresponds to the correlator:
\bea \left< \mu (z_1,\bar{z}_1) \mu (z_2,\bar{z}_2) \Theta^+ (z_3,\bar{z}_3)
\Theta^- (z_4,\bar{z}_4) \right> =  |z_{12}|^{1/2} \left( \arctan( \sqrt{z-1}) +
\overline{\arctan( \sqrt{z-1})} \right)  \nonumber \eea
where $\Omega(z,\bar{z})$ is the normal vacuum and $\Theta^{\pm}$
are a specific case at $h=0$ of the more general non-chiral
fermionic operators that we have already discussed.

For any value of $k$ we can again solve to find the conformal blocks. They are:
\bea
F_1(z)&=&  {}_2F_1 \left( \f{2}{k+2}, \f{k+1}{k+2};\f{2k+2}{k+2};z \right) \\
F_2(z)&=& z^{\f{-k}{k+2}}  {}_2F_1 \left( \f{1}{k+2},\f{2-k}{k+2};\f{2}{k+2};z
\right) \eea
Again for generic values of $k$ we must have the diagonal correlator:
\bea \label{eqn:newcorrel} G(z,\bar{z})= U_{2,2} \left\{ \f{U_{1,1}}{U_{2,2}}
|F_1|^2 + |F_2|^2 \right\} \eea
Now imposing monodromy around $z=1$ we find:
\bea \f{U_{1,1}}{U_{2,2}}= - \f{2}{ \pi^2 k^2} -\f{2}{\pi^2 k} + O(1) \eea
We therefore see that, in order to have a well defined limit in
(\ref{eqn:newcorrel}), we must take $U_{2,2} \sim k^2$ and we then find:
\bea G(z,\bar{z}) \rightarrow |\CF_1(z)|^2\nonumber \eea
Therefore we see that in the limit of the correlators we do \emph{not} find the
second solution $\CF_2(z)$ corresponding to operators $\Theta^{\pm}$. We now have a
rather interesting puzzle. For $k \ne 0$ we have no degeneracy and get a unique
correlator. However at the point $k=0$ we have a \emph{choice} of two different
correlators coming from the extra indicial structure of $\Theta^{\pm}(z,\bar{z})$.
The fundamental reason for this is that the moduli space of solutions to the
monodromy constraints:
\bea U_{1,2}e^{2 \pi i \f{k}{k+2}}= U_{1,2}\nonumber \eea
is not smooth as a function of $k$. We see in particular that the
condition is trivial if $\f{k}{k+2} \in Z \Leftrightarrow
h_{1,3}-h_{1,1} \in Z$. It is exactly in the cases in which
conformal dimensions differ by integers, and we may get
logarithms, that the monodromy constraints break down.

This conclusion is applicable to any conformal field theory in which one has an
extended multiplet structure at a certain point. The limit of the correlators is not
the same as solving the theory at the limiting  point. It would be particularly
interesting to analyse this in the context of disordered systems which can be
studied in the replica limit or using the super-symmetric
approach\cite{Bhaseen:2000mi}
\section{Conclusion}
We have investigated the structure of the $h_{1,s}$ fields in the
$c_{p,q}$ models by directly studying their correlation functions.
We found that the vacuum null vector of the irreducible theory can
be accompanied by two extra primary non-chiral fermionic fields.
We also found a chiral triplet algebra generated by $h_{1,4p-1}$
fields. For $pq \in 2Z$ we also found extra chiral fermionic
structure.

We were not able to understand the appearance of this structure but in the case of
the non-chiral partners to the vacuum null vector it naively comes from the fact:
\bea \ln \left| \f{1}{z} \right| = - \ln |z|\nonumber \eea
It is the minus sign which indicates that operators behave in a fermionic manner
under crossing symmetries. We have also seen that such an extended symmetry of
particular fields does not arise in the limit of correlation functions.

It would be interesting to understand many of these points more
thoroughly.

\section{Acknowledgements}
I am grateful to I. I. Kogan for useful and stimulating discussions. I have received
funding from the Martin Senior Scholarship awarded by Worcester College, Oxford.




\end{document}